\documentclass[prb,twocolumn,showpacs,preprintnumbers]{revtex4}

\usepackage{amsmath}
\usepackage{amssymb}
\usepackage[T1]{fontenc}
\usepackage{graphicx}

\newcommand{\eps}{\varepsilon}

\begin{document}

\title{Magnetic edge states of impenetrable stripe}
\author{A.~Matulis}
\email{amatulis@takas.lt}
\author{T.~Pyragien\.{e}}
\affiliation{Semiconductor Physics Institute, Go\v{s}tauto 11, 2600 Vilnius,
Lithuania}
\date{\today}

\begin{abstract}
The electron motion in a strong perpendicular magnetic field close
to the impenetrable stripe is considered by making use of the singular
integral equations technique. The energy spectrum is calculated
and compared with the energy spectrum of the round antidot.
It is shown that in the case of the long stripe the eigenfunctions
can be obtained as a superposition of magnetic edge modes,
while fractal energy levels obtained in a high energy region can be
explained from the quasi-classical point of view.
\end{abstract}


\maketitle

\section{INTRODUCTION}

Progress in a nanometer technology and the ability to tailor
potentials has triggered a broad activity in low-dimensional
semiconductor nanostructures. Among the 2D (two-dimensional)
structures the quantum dots with the electrons confined in a small
region have been a subject of intense theoretical and experimental
research during last years. \cite{jacak98} The complete
confinement and the discrete energy spectrum converted these
objects into a useful instrument for the electron interaction and
correlation studies.\cite{maksym00} In the strong perpendicular
magnetic field the quantum antidot, the region with a repulsive
potential, can bound the electrons as well. The magnetotransport
experiments on the arrays of the quantum antidots \cite{klitz91}
showed the close relation of the pronounced structure in the
magnetoresistance and the periodic classical orbits, or the
corresponding spectrum of the antidots. It was tested on the
arrays of various shape antidots.\cite{luth97} The importance of
the antidot-bounded electron states was confirmed in the studies
of magnetotransport through clusters of the antidots
\cite{kirtcz97} and the individual antidots.\cite{kircz94}

The spectrum of the quantum antidot in the
magnetic field is also interesting
from the dynamic chaos point of view
(see review article\cite{smy02} and references there).
The quantum antidots together with quantum billiards are the
most simple and convenient structures for revealing the links
between the auto-correlations in quantum spectrum
and the periodic orbits of the classical problem.
The most convenient technique for solving the antidot eigenvalue
problems with not separating variables is the singular integral
equations. Usually the sharp antidot edges increase the singularity
of these equations making them rather complicated and even
not useful. In this paper using the simplest antidot, the finite
impenetrable line, we demonstrate how the integral equation
technique can be used in the case of the antidot with the sharp
edges. We have failed to find that this simple but revealing antidot
spectrum has been considered ever before. Comparing this spectrum
with the round antidot one we demonstrate the main features of
the problem with nonseparable variables. The limit cases of a long
antidot and high electron energy show the peculiarities of the
quantized magnetic edge modes and the quasi-classical quantization.

The paper is organized as follows. In Sec.\ II the problem is
formulated. In Sec.\ III the simplest ultra short stripe case is
considered, and in Sec.\ IV the numerical results for the energy
spectrum are presented. In order to explain the physical meaning
of the spectrum peculiarities two limit cases --- the long stripe
in Sec.\ V and large electron energy in Sec.\ VI --- are considered.
In the last Sec.\ VII the conclusions are given, and in the
Appendix the details related to the discretization of the
singular integral equation are collected.

\section{Model}

\begin{figure}
\includegraphics[width=5cm]{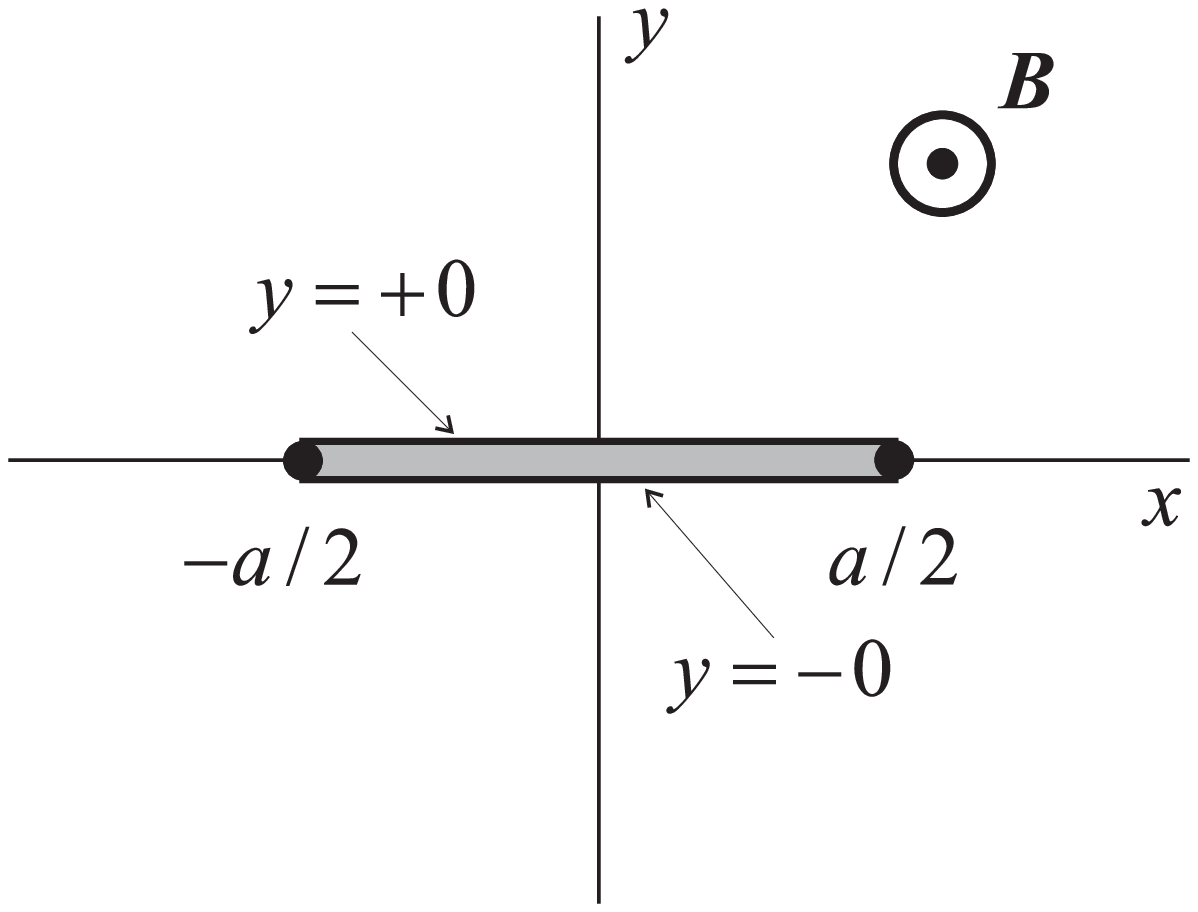}
\caption{Layout.}
\label{fig1}
\end{figure}

We consider the electron moving in $xy$-plane which is shown in Fig.\
\ref{fig1}.
The antidot, an infinitely thin impenetrable line $|x|\leqslant a/2$, $y=0$,
is indicated by the grey stripe.
We solve the Schr\"{o}dinger equation
\begin{equation}\label{schred}
  \{H-E\}\Psi({\bf r}) = 0
\end{equation}
with the following dimensionless Hamiltonian:
\begin{equation}\label{ham}
  H = - \frac{1}{2}\{\nabla+i{\bf A}({\bf r})\}^2,
\end{equation}
where the perpendicular magnetic field is described by the vector
potential in the symmetric gauge ${\bf A}({\bf r})=\{-y,x\}/2$.
We use the following notation for 2D vectors ${\bf r}=\{x,y\}$.
The energy is measured in $\hbar\omega_c$ ($\omega_c=eB/mc$) units,
and the coordinates --- in the magnetic length $l_B=\sqrt{c\hbar/eB}$
units. The antidot --- the impenetrable stripe --- is taken into account by
hard wall boundary condition
\begin{equation}\label{bc}
  \Phi({\bf r})\Big|_{|x|\leqslant a/2,y=\pm 0}=0.
\end{equation}
Besides, the wave function satisfies zero boundary conditions
at the infinity $\Psi({\bf r})|_{r\to\infty}=0$.

Making use of the Green theorem the above two-dimensional problem
can be transformed into one-dimensional integral equation.
Indeed, introducing the whole plane Green function as a solution
of the equation
\begin{equation}\label{equgreen}
  \{H-E\}G({\bf r}|{\bf r}') = -\delta({\bf r}-{\bf r}')
\end{equation}
and taking the boundary conditions $G(\infty|{\bf r}')=0$ into account,
one can present the wave function as the integral
(see the details in Ref.~\onlinecite{smy02})
\begin{equation}\label{wfint}
  \Psi({\bf r}) = \frac{1}{2}\int_{-a/2}^{a/2}dx'G({\bf r}|x',0)
  F(x')
\end{equation}
over the perimeter of the antidot. Here
\begin{equation}\label{pf}
  F(x) = \Psi_y(x,+0)-\Psi_y(x,-0)
\end{equation}
is the difference of the wave function derivatives on the opposite
sides of the stripe. We shall refer to it as the perimeter function.

The wave function defined via Eq.~(\ref{wfint}) satisfies already
equation (\ref{schred}) and the boundary condition at the infinity.
Satisfying boundary condition on the stripe (\ref{bc}) we get the
following integral equation:
\begin{equation}\label{ieq}
  \int_{-a/2}^{a/2}dx'K(x,x')F(x') = 0
\end{equation}
with the kernel
\begin{equation}\label{kernel}
  K(x,x') = 2\pi G(x,0|x',0).
\end{equation}
The nonessential factor $2\pi$ is included for the sake of
convenience.

This integral equation is our main instrument. Note that taking the derivative
of Eq.~(\ref{wfint}) over $y$ and equating it to the perimeter function
on the stripe, one more integral equation can be obtained. In our case
it isn't necessary, because our stripe-antidot has no inner region, and
consequently, there are no spurious eigenstates, which have to be
properly eliminated in the case of other antidots.\cite{horn00}

In order to fix the kernel we have to solve Green function equation
(\ref{equgreen}). The solution of it is known\cite{smy02}
\begin{subequations}\label{gf}
\begin{eqnarray}
  G({\bf r}|{\bf r}') &=& -\frac{1}{2\pi}
  \exp\left\{i[{\bf r}\times{\bf r}']_z/2\right\}g(s), \\
  g(s) &=& -\Gamma(-\eps)\exp(-s/2)U(-\eps|1|s), \\
  s &=& |{\bf r}-{\bf r}'|^2/2, \\
  \eps &=& E-1/2.
\end{eqnarray}
\end{subequations}
Here the symbol $\Gamma(z)$ stands for $\Gamma$-function, and
$U(a|b|z)$ is the Kummer function of the second kind --- the solution of the
confluent hypergeometric equation.\cite{abram64}

Inserting the above expression into Eq.~(\ref{kernel}) we
obtain the following kernel
\begin{equation}\label{kern}
  K(x,x') = -g(s), \quad s = \frac{1}{2}(x-x')^2.
\end{equation}

\section{Short stripe}

When solving integral equation (\ref{ieq}) numerically the main
problem is the kernel singularity at $x=x'$. One can expect that it
will lead to the singularity of the perimeter function at the ends
of the stripe $x=\pm a/2$. In order to reveal the above singularity
we considered the limit case of the ultra short stripe ($a\to 0$)
when the Kummer function can be replaced by
its following expansion:
\begin{equation}\label{kummexp}
  \lim_{s\to 0}U(-\eps|1|s) = -\frac{1}{\Gamma(-\eps)}\left\{
  \ln s + \psi(-\eps)-2\psi(1)\right\}.
\end{equation}
Here the symbol $\psi(z)$ stands for the logarithmic $\Gamma$-function
derivative, or the so called $\psi$-function. Note apart the singular
logarithmic term we included terms which are large close to the first
Landau level ($\eps\to 0$), and the constant term $\psi(1)=-\gamma$
($\gamma \approx 0.5772$), which we need to get the proper behavior
of the lowest antidot energy branch.

Now inserting the above expansion into Eqs.~(\ref{kern}), then into
(\ref{ieq}), and scaling the variables $x\to ax$, we arrive at the following
approximate integral equation for the ultra short stripe case:
\begin{equation}\label{ieqsh}
  \int_{-1/2}^{1/2}dx'\ln|x-x'|F(x') = \lambda \int_{-1/2}^{1/2}F(x')dx',
\end{equation}
where
\begin{equation}\label{lambda}
  \lambda = -\frac{1}{2}\left\{\ln(a^2/2)+\psi(-\eps)+2\gamma\right\}.
\end{equation}
Thus, the short stripe energy spectrum problem is reduced to the
calculation of the eigenvalues of integral equation (\ref{ieqsh}).
Then the energy can be obtained by solving algebraic equation (\ref{lambda}).
For instance, replacing the $\psi$-function by its simplest expansion in
the vicinity of the first Landau level $\psi(-\eps)\approx 1/\eps-\gamma$
we get
\begin{equation}\label{ensh}
  E-1/2=\eps = \frac{1}{\gamma-2\lambda-\ln(a^2/2)}.
\end{equation}

\begin{figure}
\includegraphics[width=7cm]{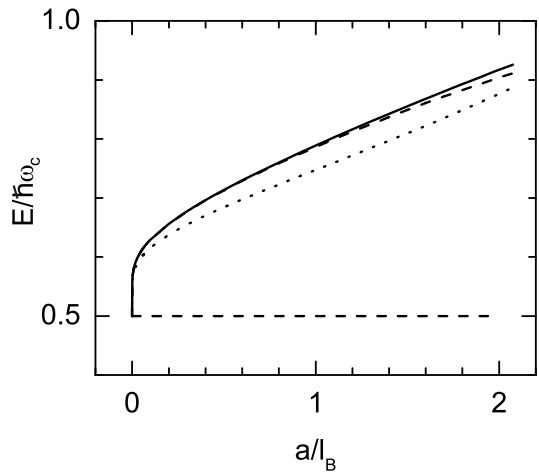}
\caption{Energy spectrum of ultra short stripe. Solid curve --
solution of Eq.~(\ref{lambda}), dotted curve -- simplified asymptotics
according to Eq.~(\ref{ensh}), and dashed curve -- numerical integration of
integral equation (\ref{ieq}).}
\label{fig2}
\end{figure}

Due to the average of the eigenfunction over the stripe
on the right-hand side of equation
(\ref{ieqsh}) it has the single not equal to zero eigenvalue $\lambda$.
It can be checked by the straightforward integration that the function
\begin{equation}\label{pfsh}
  F(x) = 1/\sqrt{1/4-x^2}
\end{equation}
satisfies integral equation (\ref{ieqsh}) with $\lambda=-2\ln 2$.
The corresponding energy spectrum branch is shown in Fig.\
\ref{fig2} as a function of the stripe length $a$: the solid curve
is obtained by the numerical solution of Eq.~(\ref{lambda}), while
the dotted curve indicates the simplified version of the
asymptotics according to Eq.~(\ref{ensh}).
Both of them coincide in the limit case $a\to 0$.
We see that the antidot (short impenetrable stripe) expels
a single level from the first degenerate Landau state (indicated
by a thick dashed horizontal line). The longer
the stripe is, the higher is the level. Note a rather fast energy grow
at small $a$ values. It is a characteristic feature of the energy
level corresponding to the non-perturbed level with zero orbital
momentum, which has a non zero electron density at the origin
${\bf r}=0$. All other non-perturbed levels have zero electron density there,
and thus, they are weakly influenced by the stripe, and consequently,
not expelled in this simplest ultra short stripe approach.

\section{Numerical results}
\label{results}

The most important result for the ultra short stripe case presented
in the previous Section is perimeter function (\ref{pfsh}) which
singularity at the stripe ends is caused by the interplay of
the logarithmic singularity of the kernel and the sharp antidot
edges. Thus, it is inherent to the perimeter function of
general integral equation (\ref{ieq}) as well.
That is why in order to achieve the proper accuracy in numerical
solution of the above equation one has to take both singularities
(kernel and perimeter function) into account explicitly.
For this purpose we have replaced the perimeter function as follows:
\begin{equation}\label{pfn}
  F(x) = \frac{f(x)}{\sqrt{a^2/4-x^2}},
\end{equation}
and discretized the obtained integral equation for the function
$f(x)$ including the singular factors into the proper weights of
the discretization scheme. See
the details of the calculation in Appendix \ref{details}.
Instead of solving the obtained matrix equation
\begin{equation}\label{matrequ}
  \mathcal{K}{\bf f} = 0
\end{equation}
the corresponding eigenvalue problem
\begin{equation}\label{eigen}
  \mathcal{K}{\bf f}_n = \lambda_n{\bf f}_n
\end{equation}
for various electron energies $\eps$ was considered. The electron
energy was defined by zeroing the obtained eigenvalues
$\lambda_n=\lambda_n(\eps)=0$.

\begin{figure}
\includegraphics[width=7.5cm]{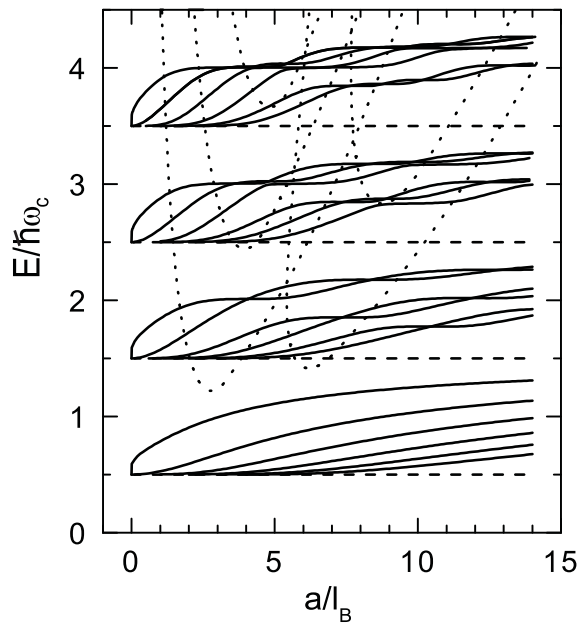}
\caption{Energy spectrum of the stripe.
The non-perturbed Landau levels are indicated by thick dashed
horizontal lines. The solutions of Eqs.~(\ref{equnumb})
and (\ref{equnumb3}) are indicated by dotted curves.}
\label{fig3}
\end{figure}

The obtained stripe energy spectrum is shown in Fig.\ \ref{fig3}
where six highest levels expelled from each Landau level (dashed
horizontal lines) are indicated.
On the axes the original dimensions are shown. Thus, not only
the energy dependence on the stripe length but its dependence on the
magnetic field strength ($a/l_B\sim\sqrt{B}$) can be traced as well.

\begin{figure}
\includegraphics[width=7.5cm]{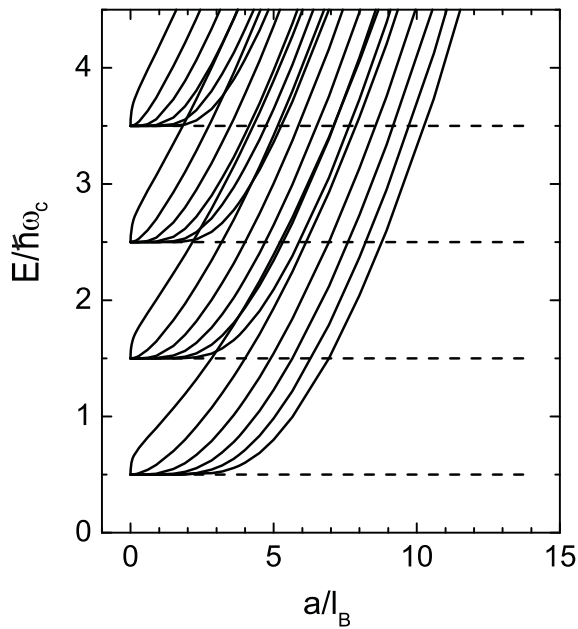}
\caption{Energy spectrum of the round antidot.}
\label{fig4}
\end{figure}

For a comparison the spectrum of round impenetrable antidot with
a diameter $a$ is shown in Fig.\ \ref{fig4}.
It was obtained by means of zeroing the radial antidot
wave function (it coincides with Green function
(\ref{gf}) with ${\bf r}'=0$ assumed) at the antidot border $r=a/2$.

The characteristic feature of both spectra is fast expelled
first antidot level for each Landau state in the case of small
$a$ values. The detailed behavior of this level expelled from
the first Landau level for the stripe is shown in Fig.~\ref{fig2}
by a dashed curve. Rather good coincidence of it with the short
stripe energy (solid curve) confirms a good accuracy of the developed
numerical scheme.

We see that these two spectra of the antidot-stripe and the round antidot
differ essentially. The round antidot spectrum is a typical one
for the system with separable variables. In this case the variables can be
separated due to the cylindric symmetry of the problem, and actually we
have independent radial problems for every angular momentum value,
which energy spectrum branches freely intersect each other. The main point
is that when the antidot level with some orbital momentum $l$ reaches the next
Landau level, the antidot level with the same momentum $l$ is already expelled
from it. Consequently, any antidot level freely crosses any Landau level.
And we see the energy spectrum branches going up when $a$ increases
with numerous crossings.

This is not the case for a stripe spectrum. Due to the lack of symmetry
the orbital momentum is not good quantum number any more, and instead
of crossings we have anti-crossings. Moreover, expelled antidot levels
can not cross Landau levels any more. So, when the parameter $a/l_B$ grows
the expelled antidot levels saturate bellow the next Landau level.

Nevertheless there are still some crossings. See, for instance, the
behavior of levels expelled from the second Landau level in Fig.\ \ref{fig3}.
The matter is that the stripe in the perpendicular magnetic field still
has the inversion symmetry (${\bf r}\to -{\bf r}$). Due to it all the perimeter
functions (and the wave functions as well) can be divided into the symmetric
and the anti-symmetric ones which actually satisfy the different integral
equations. Thus, in the energy spectrum of the stripe crossings between
the symmetric and anti-symmetric spectrum branches are possible.
These crossings and the occurring waviness of the spectrum branches
are the most prominent feature of the considered stripe spectrum.
Note that when the electron energy grows the above waviness is transformed
into numerous pronounced plateaux on the spectrum branches. Now we are
going to explain the physical meaning of these plateaux and waviness
considering two limit cases of the long stripe and large electron energies.

\section{Long stripe}

\begin{figure}
\includegraphics[width=7.5cm]{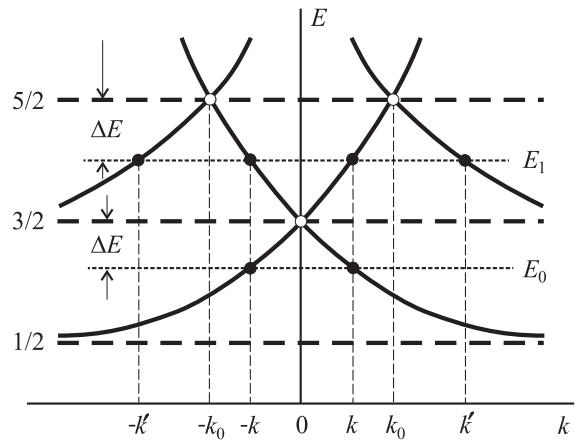}
\caption{Spectrum of the magnetic edge modes.}
\label{fig5}
\end{figure}

Let us start with the long stripe approximation. When the stripe is
infinite there are two types of magnetic edge modes propagating
along the stripe on both its sides. In this case we have the
problem invariant under the translation along the stripe, and thus,
the corresponding eigenvalues can be labelled by the electron
momentum component $k$ along $x$-direction.
The eigenfunctions can be expressed in
terms of parabolic cylinder function as $\Psi(k|{\bf r})=
\exp(\pm ikx)D_{\eps}\left\{\pm \sqrt{2}(y-k)\right\}$.\cite{levinson91}
The spectrum is obtained zeroing the wave function on the stripe
$D_{\eps}(\mp\sqrt{2}k)=0$, and it is shown in Fig.\ \ref{fig5}.
We see that it consists of two independent systems of branches describing
the electron motion to the left above the stripe, and to the right ---
below it. In the asymptotic region $k\to\pm\infty$ the branches
tend to the Landau levels shown by thick dashed horizontal lines.
The intersection points of the above branches coincide with the Landau
levels as well.

In the case of the finite stripe the electron motions above and under
the stripe are no more independent, because moving above the stripe
the electron reaches its end, bends around the corner, continues
its motion under the stripe, and so on. Bending of the corner is
a rather complicated diffraction problem, but in the asymptotic long
stripe case we can replace it by some scattering matrix acting on
the longitudinal motion exponents.

The description of electron motion depends on the number of edge modes
participating in it. For instance, if the electron energy is lower
than the second Landau level with energy $3/2$ (see, the lower thin dotted
horizontal line in Fig.\ \ref{fig5} labelled by $E_0$) there are only two
edge modes indicated by solid circles.
One of them with momentum $k$ moves above the stripe to the left,
while the other one with momentum $-k$ moves to the right under it.
At the stripe ends these edge modes are scattered one into another.
As there is a single scattering channel only, the scattering
probability is equal to unity. Consequently, due to the scattering event
the electron wave function is multiplied by some scattering amplitude
$S=\exp\{i\chi(\Delta E)\}$, while the propagation along the stripe
can be taken into account by the propagator $\exp(ika)$. Thus,
taking into account the periodic motion of the electron (after
bending both stripe ends the exponential part of the electron wave function
must coincide with itself), we can write
down the following simple rule for quantization of the edge modes
in the asymptotic long stripe case: $\exp\{2i(ka+\chi)\} = 1$, or
\begin{equation}\label{asquant}
  ka + \chi(\Delta E) = \pi n, \quad n = 1,2,\cdots.
\end{equation}

Now using the relation $\Delta E = vk$ ($v = 5/4\sqrt{2}$) which
follows from the properties of the parabolic cylinder functions close
to the intersection point at $E=3/2$, $k=0$, and the expansion
\begin{equation}\label{chiexp}
  \chi(\Delta E) = \chi_0 + \chi_1\Delta E,
\end{equation}
we solve Eq.~(\ref{asquant}) and get
\begin{equation}\label{Eas1}
  \Delta E_n = \frac{A(n-\Delta)}{a + \delta}, \quad
  A = \frac{4\pi\sqrt{2}}{5} \approx 3.55.
\end{equation}
Two other parameters: the effective elongation of the
stripe $\delta = v\chi_1$ and the quantum number defect $\Delta = \chi_0/\pi$
depend on the scattering amplitude phase and unfortunately, can not
be found analytically.

Fitting the numerically obtained energy branches expelled from
the first Landau level in the interval $50<a<100$ by $E=A_n/(a+\delta_n)$
we have obtained the following parameters:
\begin{center}
\begin{tabular}{c||c|c|c|c|c|c}
  $n$        & 1    & 2    & 3     & 4 & 5 & 6  \\\hline
  $\delta_n$ & 4.29 & 5.53 & 5.69  & 7.25  & 7.99 & 9.49 \\
  $A_n$      & 3.48 & 7.04 & 10.40 & 14.01 & 17.39 & 20.99 \\
  $A_n/A$    & 0.99 & 1.98 & 2.92  &  3.93 & 4.89  & 5.88
\end{tabular}
\end{center}

\noindent
Note the numbers in the last row coincide rather well with the
integers $n$ what convince us that the picture of quantized edge modes
is quite adequate.

To explain the behavior of energy branches expelled from the upper Landau
levels is more complicated because there are more edge modes
present. For instance, close to the third Landau level (see
the upper thin dotted horizontal line in Fig.\ \ref{fig5} labelled
by $E_1$) there are four edge modes. Two of them with momenta $-k$
and $k'$ propagate above the stripe to the left, while the other
two with momenta $k$ and $-k'$ propagate under it to the right.
Consequently, in this case the propagation of electron on both sides
of the stripe has to be described by the following propagator:
\begin{equation}\label{prop}
  \mathcal{P}(x) = \begin{pmatrix} e^{-ikx} & 0 \\
  0 & e^{ik'x} \end{pmatrix}
\end{equation}
acting on the state vector
\begin{equation}\label{stvec}
  \Psi = \begin{pmatrix} A \\ B \end{pmatrix}
\end{equation}
composed of edge mode superposition coefficients.
Bending of the edges is characterized by some $2\times 2$ scattering
matrix $\mathcal{S}$.
Now the quantization is performed by the following self-consistency
condition:
\begin{equation}\label{scc}
  \mathcal{SP}(a)\mathcal{S}\mathcal{P}(a)\Psi = \Psi,
\end{equation}
and the energy spectrum of the stripe can be defined
zeroing the determinant of the above equation
\begin{equation}\label{sccdet}
  \det|\mathcal{SP}(a)\mathcal{S}\mathcal{P}(a)-\mathcal{I}|=0.
\end{equation}

The absolute values of the scattering matrix $\mathcal{S}$ elements are
given in Ref.\ \onlinecite{levinson91}. We see that close to the third
Landau level the absolute value of the off-diagonal elements are nearly unity
($|S_{01}|=|S_{10}|\approx 1$), while the diagonal elements are small
$|S_{00}|=|S_{11}|\approx \Delta k/2k_0$ where the symbol $k_0$
stands for the edge mode intersection point and $\Delta k = k'-k$ is
the difference of edge mode momenta, propagating on both sides of the stripe.
Thus, adding the phases we construct the following scattering matrix
\begin{equation}\label{scattmatr}
  \mathcal{S} = e^{i\Phi}\begin{pmatrix}
  e^{i\varphi}\Delta k/2k_0 & e^{i\psi} \\ e^{-i\psi} &
  -e^{-i\varphi}\Delta k/2k_0e^{i\varphi}
  \end{pmatrix}
\end{equation}
which satisfies the unitarity condition $\mathcal{SS}^+=\mathcal{I}$
with the accuracy of $\Delta k$ terms.

Now expanding the momenta close to the intersection points as
\begin{subequations}\label{expmom}
\begin{eqnarray}
&&  k = k_0-\frac{1}{v_1}\Delta E, \quad
  k' = k_0+\frac{1}{v_2}\Delta E, \\
&&  k' - k = \frac{2}{v'}\Delta E, \quad
  v' = \frac{1}{2}\left(\frac{1}{v_1} + \frac{1}{v_2}\right)^{-1},
\end{eqnarray}
\end{subequations}
we transform Eq.~(\ref{sccdet}) into the following approximate equation
\begin{equation}\label{twocaneq}
  \exp(ia\Delta E/v') - \exp(-2i\Phi) = \pm
  \frac{i\Delta E}{k_0v'}\sin(k_0a-\varphi).
\end{equation}
If one neglects the small term on the right-hand side of this
equation one gets the result similar to Eq.~(\ref{Eas1})
\begin{equation}\label{twoen0}
  \Delta E_n^{(0)} = \frac{A'(n-\Delta')}{a + \delta'},
\end{equation}
where $\Phi(\Delta E)= \Phi_0+\Phi_1\Delta E$, $A'=\pi v$,
$\Delta'=\Phi_0/\pi$ and $\delta'=\Phi_1v$.
The right-hand side term taken into account as a perturbation leads
to the following oscillating correction:
\begin{equation}\label{twoen1}
  \Delta E_n^{(1)} = \pm \frac{\Delta E_n^{(0)}}{k_0a}\sin(k_0a-\varphi).
\end{equation}
It is remarkable that the period of oscillations depends on
$k_0=1/\sqrt{2}$ only. It leads to $\Delta a=2\pi/k_0\approx 8.89$,
what coincides rather well with the period value $8.8$ obtained from
the numerical calculation result for the two upper levels expelled from
the second Landau level.

\begin{figure}
\includegraphics[width=6cm]{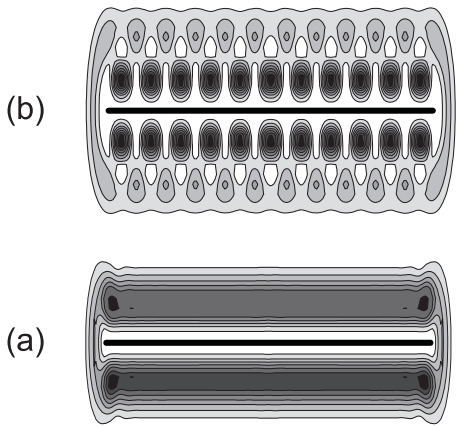}
\caption{Electron density for long stripe ($a/l_B=48$): (a) -- for
the highest antidot
level expelled from the first Landau level, and (b) -- for the same
level expelled from the
second Landau level.}
\label{fig6}
\end{figure}

This simple asymptotic picture of interfering magnetic edge modes is
confirmed by Fig.\ \ref{fig6} where the contour plots of the
electron densities corresponding
to the above considered antidot states are shown.
We see that the wave function for the antidot level expelled from the first
Landau level looks like a cigar (and it does not matter how long
is the stripe) what indicates that it is composed of a single magnetic
edge mode. In the case of the antidot level expelled from the second Landau
level there are the lumps (the longer the stripe is the more lumps there are)
which are caused by the interference of two pairs of
edge modes propagating on both sides of the stripe.

\section{Quasi-classical limit}

\begin{figure}
\includegraphics[width=7.5cm]{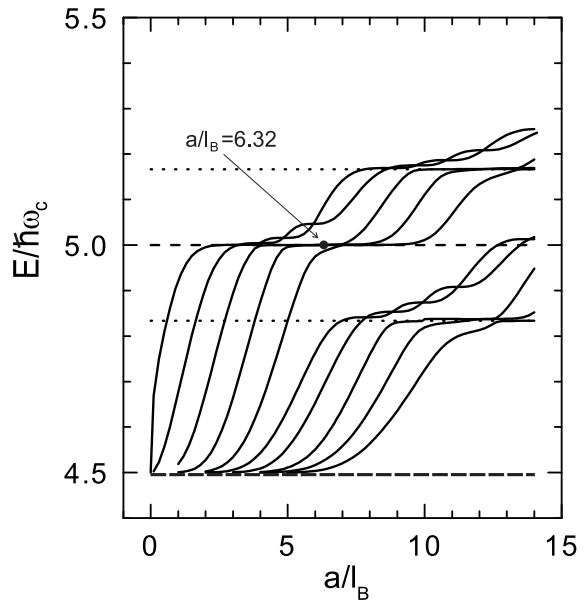}
\caption{The stripe energy levels expelled from the fifth Landau level.}
\label{fig7}
\end{figure}

As it has already been mentioned in Sec.\ \ref{results} (see Fig.\ \ref{fig3})
the oscillations surveyed close to the lowest Landau levels
change into well pronounced plateaux when the number of Landau level
is incremented. The detailed view for the antidot levels expelled
from the fifth Landau level is shown in Fig.\ \ref{fig7}.
It is remarkable that the energy of these plateaux is very close to
simple fractions of cyclotron energy. Thus, the plateau indicated by a dashed
horizontal line is right in the middle between two adjacent Landau
levels with energy $E=9/2+1/2=5$, while the energy of two other plateaux
indicated by dotted horizontal lines exceed the fifth Landau level energy
($9/2$) by one and two thirds.

Unfortunately, in the case of high Landau levels the number of the interfering
edge modes is large, and this fact makes it difficult to apply the long stripe
approximation considered in the previous section. Nevertheless the
simplified description is still possible due to the large electron
energy.

It is known that when the electron energy is large the
quasi-classical approach based on Bohr quantization rule
can be used. In the case
of free 2D electron in the homogeneous perpendicular magnetic
field it reduces to the estimation of the following integral:
\begin{equation}\label{phint}
  \oint \xi d\eta = 2\pi n + \pi, \quad n = 0,1,2,\cdots
\end{equation}
composed of fast coordinates \cite{matulis02}
\begin{equation}\label{fc}
  \xi = p_x-y/2, \quad \eta = p_y + x/2
\end{equation}
over the electron trajectory.
Inserting the solution $\xi = \sqrt{2E}\cos t$, $\eta = \sqrt{2E}\sin t$
into Eq.~(\ref{phint}) one immediately gets a well known expression for
Landau level energy $E_n=n+1/2$.
Note the considered electron motion is two-dimensional, and consequently,
two more coordinates --- the slow motion coordinates
\begin{equation}\label{slc}
  X = x/2 - p_y, \quad Y = y/2 + p_x
\end{equation}
--- have to be taken into account. In free electron case it is trivial
because the Hamiltonian does not depend on them. The single important
thing is the commutator $[X,Y]=i$, which shows that the total slow
coordinate phase volume divided by $2\pi$ gives the degeneracy of the
corresponding energy level.

\begin{figure}
\includegraphics[width=6cm]{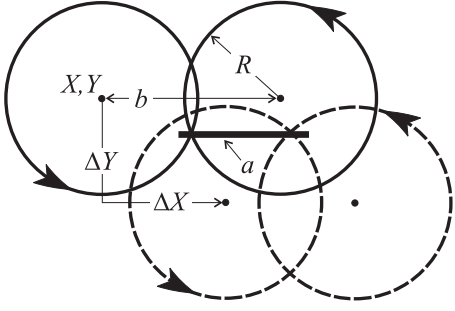}
\caption{Two circle trajectory.}
\label{fig8}
\end{figure}

The influence of the stripe on the classical electron trajectory
can be taken into account via scattering events. The simplest trajectory
with two scattering events is shown in Fig.\ \ref{fig8} by a solid curve.
We see that in this case the trajectory is composed of two
Larmor circles, and consequently, it is twice longer than the trajectory
of the free electron rotating in the magnetic field. Thus, the integral
on the left-hand side of Eq.~(\ref{phint}) becomes twice larger. This fact
leads to the twice smaller separation of energy levels ($\Delta E = 1/2$)
as compared with the separation of Landau levels.

In order to decide whether the fractal Landau levels obtained in this
quasi-classical way can take place or not, one has to inspect the
slow motion coordinates. We shall take them into account in the most
simple way. Note that there are more equivalent trajectories
with two scattering events and the same energy (the same radius of
the Larmor circle). Let us mark them by the position of the left circle
center ($X,Y$). So, changing the center by some ($\Delta X,\Delta Y$)
we obtain another equivalent trajectory, as it is shown
in Fig.\ \ref{fig8} by a dashed curve.
Thus, the integral over all possible coordinates $X$ and $Y$
\begin{subequations}\label{phvol2}
\begin{eqnarray}
&&  V_2(R,a) = \int\int dXdY \nonumber \\
&&  = \int dY n_2(R,2\sqrt{R^2-Y^2}), \\
&&  n_2(R,b) = (2b-a)\Theta(2b-a)\Theta(a-b) \nonumber \\
&& \phantom{mmmmm} + a\Theta(b-a)
\end{eqnarray}
\end{subequations}
gives the total phase volume for the trajectories with given radius
$R$. Here $b$ is the distance between the centers of both circles.
We assume that the quantity $N_2(E,a) = V_2(R,a)/2\pi$ gives the
degeneracy of the quasi-classical antidot level with given energy
$E=R^2/2$. The integral in Eq.~(\ref{phvol2}) can be easily calculated,
and it leads to the following number of degeneracy of quantum level
corresponding to the classical two-circle trajectory:
\begin{subequations}\label{ndeg2}
\begin{eqnarray}
&&  N_2(E,a) = \frac{2E}{\pi}\rho_2(a/\sqrt{8E}), \\
&&  \rho_2(x) = \zeta(x)\Theta(1-x) + \frac{\pi}{2}\Theta(x-1)
  -\zeta(x/2),\phantom{mm} \\
&&  \zeta(x) = \arcsin x + x\sqrt{1-x^2}.
\end{eqnarray}
\end{subequations}
If the above number is less than unity, the level does not
manifest itself. The energy values obtained by solving equation
\begin{equation}\label{equnumb}
  N_2(E,a) = n, \quad n = 1,2,3
\end{equation}
are indicated in Fig.\ \ref{fig3} by dotted curves on its left side.
We see that the larger is the energy the longer is the fractal plateau
corresponding to the quasi-classical level, and the higher is its
degeneracy. Moreover, there is a good coincidence of dotted curves
with the plateaux ranges obtained by the numerical calculation.

In a similar way the phase volume and the degeneracy of the
quasi-classical levels corresponding to three circle classical
trajectories can be estimated. In this case one can obtain
\begin{subequations}\label{ndeg3}
\begin{eqnarray}
&&  n_3(R,b) = \Theta(2R-b)\Theta(a-b)\Theta(3b-a) \nonumber \\
&& \phantom{m}\times{(a-b)\Theta(2b-a)+(3b-a)\Theta(a-2b)}, \\
&&  N_3(E,a) = \frac{E}{3\pi}\rho_3(a/\sqrt{8E}), \\
&&  \rho_3(x) = \left\{ \begin{array}{ll} 8\zeta(x/2)
  -6\zeta(x/3) - 2\zeta(x), & 0<x<1; \\
  8\zeta(x/2) - 6\zeta(x/3) - \pi, & 1<x<2; \\
  3\pi - 6\zeta(x/3), & 2<x<r.
  \end{array}\right. \nonumber \\
\end{eqnarray}
\end{subequations}
The energy values obtained by solving equation
\begin{equation}\label{equnumb3}
  N_3(E,a) = n, \quad n=1,2
\end{equation}
are shown by dotted curves on the right side of
Fig.\ \ref{fig3} as well. We see that they also correlate well
with the ranges of plateaux exceeding by $1/3$ and $2/3$ the
corresponding Landau levels.

\begin{figure}
\includegraphics[width=7cm]{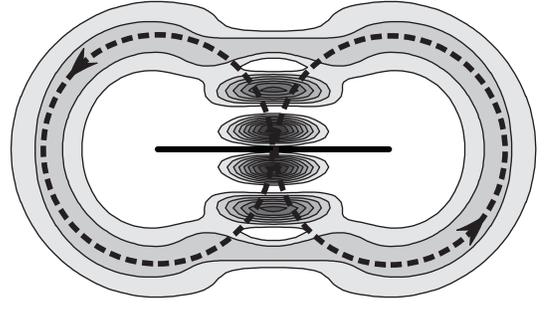}
\caption{Electron density for the level indicated by short dash in
Fig.\ \ref{fig7} corresponding to the classical two-circle
trajectory.} \label{fig9}
\end{figure}

A good agreement of all dotted curves with the degeneracy
of the fractional plateaux obtained in the numerical solution
of the problem convinces us of the adequacy of considered
quasi-classical quantization scheme. The above picture is confirmed
by the electron density plot presented in Fig.\ \ref{fig9}.
The density is calculated for the third antidot level expelled from the
fifth Landau state in the case $a/l_B=6.32$ indicated by
small solid circle in Fig.\ \ref{fig7}. We see a rather good correlation
of the electron density with the classical two-circle trajectory
shown by a dotted curve in Fig.\ \ref{fig9}.

\section{Conclusions}

The energy spectrum of the electron moving in the perpendicular magnetic field
in a vicinity of impenetrable stripe and the corresponding densities are
calculated by making use the integral equation technique for the antidot
perimeter function (the perpendicular wave function derivative at the
antidot border). It is shown that the perimeter function singularities
caused by sharp edges of the antidot can be overcome by proper discretization
technique which takes explicitly into account the logarithmic singularity
of the kernel and root-type singularities of the perimeter function.

The antidot in the magnetic field expels the antidot energy levels
from every degenerate Landau state. In the case of the round antidot
due to the circular symmetry expelled antidot levels go up when
the radius of the antidot increases (or the magnetic field strength
increases) and freely intersect each other
and the Landau levels. In the case of the antidot-stripe due to the
lack of the symmetry the variables can not be separated, and nearly all
crossings are replaced by the anti-crossings. The expelled antidot
levels can not cross the Landau levels, and consequently, they
saturate below the next Landau level when the stripe length
increases.

As the antidot-stripe still has the inversion symmetry only the pairs
of symmetric and anti-symmetric levels cross each other demonstrating
the characteristic oscillations of the spectrum branches expelled
from the excited Landau levels. These oscillations and the above
mentioned saturation can be explained in the asymptotic long
stripe case by the interference of the magnetic edge modes.

When the electron energy increases (for the antidot levels expelled
from the higher Landau levels) the above mentioned oscillations of
the spectrum branches is transformed into plateaux at the fractal
cyclotron energy values. It is shown that these fractal plateaux
can be explained using simple quasi-classical quantization rule,
and they are related to the classical trajectories composed of
several Larmor circles.

The above mentioned fractal plateaux of energy branches have to
be seen in magnetoresitance of arrays of stripe type antidots,
and the magnetization which is just proportional to the electron
energy derivative over the magnetic field strength (in considered
case through the dimensionless stripe length $a$).

\begin{acknowledgments}
We thank U.~Smilansky for helpful discussions.
\end{acknowledgments}

\appendix

\section{Discretization of singular integral equation}
\label{details}

In this section some details of the numerical solution of
integral equation (\ref{ieq}) are given.
For the sake of convenience we scale the variables $x\to ax$,
and rewrite separately the equation for the symmetric and anti-symmetric
perimeter function
\begin{equation}\label{sapf}
  F^{\pm}(-x) = \frac{1}{2}\{F(x) \pm F(-x)\}.
\end{equation}
Now the equation reads
\begin{equation}\label{ieqsa}
  \int_0^{1/2}dx'K^{\pm}(x,x')F^{\pm}(x') = 0
\end{equation}
with the symmetric (or anti-symmetric) kernel
\begin{equation}\label{kernelsa}
  K^{\pm}(x,x') = K(ax,ax') \pm K(ax,-ax').
\end{equation}
Next, we write down explicitly the singularity of the perimeter
function
\begin{equation}\label{singpf}
  F^{\pm}(x) = \frac{f^{\pm}(x)}{\sqrt{1/2-x}},
\end{equation}
and the logarithmic singularity of the kernel
\begin{equation}\label{singker}
  K^{\pm}(x,x') = 2\ln|x-x'| + \tilde{K}^{\pm}(x,x')
\end{equation}
where the second part of the kernel $\tilde{K}^{\pm}$ is
regular function at $x=x'$.
It enables us to rewrite the integral equation as follows:
\begin{eqnarray}\label{ieqmod}
&&  \int_0^{1/2}\frac{dx'2\ln|x-x'|f^{\pm}(x')}
  {\sqrt{(1/2-x)(1/2-x')}} \nonumber \\
&&  +\int_0^{1/2}\frac{dx'\tilde{K}^{\pm}(x,x')f^{\pm}(x')}
  {\sqrt{(1/2-x)(1/2-x')}} = 0.
\end{eqnarray}
Note both terms of the equation are divided by factor ${\sqrt{1/2-x}}$ in
order to have the symmetric kernel, as the numerical calculation
of corresponding symmetric matrix eigenvalues can be performed
with greater accuracy than those for the non symmetric one.

The discretization of Eq.~(\ref{ieqmod}) has been performed as follows.
The perimeter function has been replaced by vector
${\bf f}^{\pm} =\{f_0,f_1,\cdots\}$ with components $f_n=f^{\pm}(x_n)$,
and $x_n = h(n+1/2)$, $h=1/2N$. Integral equation (\ref{ieqmod})
itself has been rewritten in the form of the following matrix equation:
\begin{equation}\label{matreq}
  \{\mathcal{A} + \mathcal{B}^{\pm}\}{\bf f}^{\pm} = 0,
\end{equation}
where the corresponding matrix elements of the kernel are defined as
\begin{subequations}\label{matrelem}
\begin{eqnarray}
  A_{nm} &=& \int_{nh}^{(n+1)h}dx \nonumber \\
&&  \times\int_{mh}^{(m+1)h}\frac{2\ln|x-y|dy}{\sqrt{(1/2-x)(1/2-y)}}, \\
  B_{nm}^{\pm} &=& \tilde{K}^{\pm}(x,x')B_nB_m, \\
  B_n &=& \int_{nh}^{(n+1)h}\frac{dx}{\sqrt{1/2-x}}.
\end{eqnarray}
\end{subequations}
Both integrals can be calculated straightforwardly , and the analytical
expressions for the discretization weights obtained.

Using the discretized perimeter function ${\bf f}^{\pm}$ the electron
wave function (and the corresponding density) has been obtained via discretized
version of Eq.~(\ref{wfint}). Calculating the wave function outside
the stripe (on the stripe it is equal to zero) only the perimeter
function singularity has to be taken into account. Thus, Eq.~(\ref{wfint})
can be replaced by
\begin{equation}\label{wf}
  \Psi^{\pm}({\bf r}) = \frac{a}{2}\sum_{n=0}^{N-1}
  B_n\left\{G(a{\bf r}|ax',0) \pm G(a{\bf r}|-ax',0)\right\}f_n.
\end{equation}

\end{document}